# Legal Sentiment Analysis and Opinion Mining (LSAOM): Assimilating Advances in Autonomous AI Legal Reasoning

**Dr. Lance B. Eliot**
Chief AI Scientist, Techbruim; Fellow, CodeX: Stanford Center for Legal Informatics
Stanford, California, USA

**Abstract**

An expanding field of substantive interest for the theory of the law and the practice-of-law entails Legal Sentiment Analysis and Opinion Mining (LSAOM), consisting of two often intertwined phenomena and actions underlying legal discussions and narratives: (1) Sentiment Analysis (SA) for the detection of expressed or implied sentiment about a legal matter within the context of a legal milieu, and (2) Opinion Mining (OM) for the identification and illumination of explicit or implicit opinion accompaniments immersed within legal discourse. Efforts to undertake LSAOM have historically been performed by human hand and cognition, and only thinly aided in more recent times by the use of computer-based approaches. Advances in Artificial Intelligence (AI) involving especially Natural Language Processing (NLP) and Machine Learning (ML) are increasingly bolstering how automation can systematically perform either or both of Sentiment Analysis and Opinion Mining, all of which is being inexorably carried over into engagement within a legal context for improving LSAOM capabilities. This research paper examines the evolving infusion of AI into Legal Sentiment Analysis and Opinion Mining and proposes an alignment with the Levels of Autonomy (LoA) of AI Legal Reasoning (AILR), plus provides additional insights regarding AI LSAOM in its mechanizations and potential impact to the study of law and the practicing of law.

# 1 Background on Legal Sentiment Analysis and Opinion Mining

In Section 1 of this paper, the literature on Legal Sentiment Analysis and Opinion Mining (LSAOM) is introduced and addressed. Doing so establishes the groundwork for the subsequent sections. Section 2 introduces the Levels of Autonomy (LoA) of AI Legal Reasoning (AILR), which is instrumental in the discussions undertaken in Section 3. Section 3 provides an indication of the field of Legal Sentiment Analysis and Opinion Mining as applied to the LoA AILR, along with other vital facets. Section 4 provides various additional research implications and anticipated impacts upon salient practice-of-law considerations.

This paper then consists of these four sections:
- Section 1: Background on Legal Sentiment Analysis and Opinion Mining
- Section 2: Levels of Autonomy (LOA) of AI Legal Reasoning (AILR)
- Section 3: LSAOM and LoA AILR
- Section 4: Additional Considerations and Future Research

## 1.1 Overview of Legal Sentiment Analysis and Opinion Mining

An expanding field of substantive interest for the theory of the law and the practice-of-law entails Legal Sentiment Analysis and Opinion Mining (LSAOM), based to a great extent on research and studies of judicial behaviors such as those of jurors, and likewise the expressions of judges [48] [53] [77] [82].

Legal Sentiment Analysis and Opinion Mining can be applied to essentially any legal-related actors involved in the legal process and does not need to be limited to judges or jurors, being able to encompass other participants such as lawyers, paralegals, expert witnesses, and the like [17] [41] [75] [79].

Generically, Sentiment Analysis (SA) is a discipline that brings together a mixture of linguistics, human and social behaviors, psychology, cognitive science, and other fields to try and ascertain the sentiment expressed during some discourse. As stated by Babu and Rawther [5]: "Sentiment Analysis is the process of analyzing the sentiments and emotions of various people in various situations."

Legal Sentiment Analysis is SA that has been particularly tuned or customized to legal discourse and the legal realm.

More formally:

*Legal Sentiment Analysis entails the detection of expressed or implied sentiment about a legal matter within the context of a legal milieu.*

Another area of attention involves Opinion Mining (OM), which is also a generic form of analysis that can be applied to a legal-specific context. Per the work of Hemmation and Sohrabi [45]: "Opinion mining is considered as a subfield of natural language processing, information retrieval, and text mining. Opinion mining is the process of extracting human thoughts and perceptions from unstructured texts."

Legal Opinion Mining is OM that has been particularly tuned or customized to legal discourse and the legal realm.

More formally:

*Legal Opinion Mining (OM) entails the identification and illumination of explicit or implicit opinion accompaniments immersed within legal discourse.*

Likened somewhat to reading the tea leaves, as it were, the use of Sentiment Analysis in a legal context can aid in gauging the attitudes and feelings of those involved in legal discourse. Likewise, Opinion Mining can be useful in attempting to discern the nature of an opinion that is being expressed. Also, in the case of OM, there is often a keen interest in determining whether the expressed opinion appears to be fact-based or might be construed as non-factually based (this is further elucidated in Section 3).

**1.2 Separability of Legal SA and Legal OM**

In this paper, Legal Sentiment Analysis and Legal Opinion Mining are considered as separate and distinct from each other, as they are construed to be independent constructs, able to operate, or be utilized of each upon their own accord.

Imagine for discussion sake the extreme, whereby Legal Sentiment Analysis solely focuses on the emotional or feelings characterizations, while Legal Opinion Mining focuses solely on the identification of opinions. This is certainly viable and productive. Both though are admittedly and intentionally apt to be used in conjunction, at the same time and potentially each informing the other, in order to do a more substantive job of their respective tasks. An opinion might very well be wrapped within an aura of emotion and feelings. And, alternatively, emotion or feelings might be emoted via the expression of an opinion. There is a type of duality that can readily and often frequently arises between undertaking a Sentiment Analysis and that of an Opinion Mining effort.

But this does not render them inseparable and nor distinctly undistinguishable of each other.

The reason to provide such an emphasis on this matter of being either one-and-the-same or being separable is due to the conventional manner in which SA and OM are treated in (especially) the AI literature. For example, as stated in [38]: "Sentiment analysis, also known as opinion mining (OM), is defined as figuring out the public attitude of individuals toward distinct topics and news." Note that SA and OM are portrayed as synonymous rather than as differing.

Another example of this commingling includes this indication in [63]: "Sentiment analysis and opinion mining is an area that has experienced considerable growth over the last decade. This area of research attempts to determine the feelings, opinions, emotions, among other things, of people on something or someone. To do this, natural language techniques and machine learning algorithms are used." In this example, the two are not as explicitly declared as synonymous, and instead are treated as though effectively they are the same, albeit also proffering that they are perhaps named differently.

The usage here, in this paper, for clarification, assumes that (at least) <u>Legal</u> Sentiment Analysis and the reference to <u>Legal</u> Opinion Mining are each distinctive of each other, and though they are oftentimes joined at the hip, as it were, they are nonetheless still each a separate form of analysis and have different results or aims of what they seek to produce. The emphasis on the *legal incarnations* of SA and OM allows, perhaps, for the generic SA and generic OM to be considered as merged or inextricably intertwined (as per the dominance in the AI literature thereof).

One other related facet that partially explains why SA and OM are so closely coupled in the AI literature is due to the incorporation of in-common AI techniques and technologies. In that sense, since the underlying architecture or technological ecosystem is of the same ilk, it is easiest to then blend together that SA and OM are the same. The view here is that despite the possibility of the alike system underpinnings, they are still separable.

1.3 **Legal SA and Legal OM: Notable Exemplar**

As a vivid and classic demonstration of the use of sentiment and opinion in a legal context, many American attorneys are likely aware of the famous *Tribute to a Dog*, a matter that arose in a court case from the U.S. courts in the 1870s that is often taught or cited in law schools even still today. This is worthwhile to briefly explore herein, showcasing how sentiment and opinion are at times employed in a legal context.

Here are the particulars of the *Tribute to a Dog* matter.

In the case of *Burden v. Hornsby*, a case tried in Pettis County of Missouri on September 23, 1870, attorney George Graham Vest represented Charles Burden in a case against a sheep farmer named Leonidas Hornsby, accused of killing "Old Drum," a local hunting dog owned by Burden, the dog having been killed on October 18, 1869 while on Hornsby's farm.

Purportedly, Hornsby had beforehand vowed to kill any dog that wandered onto his farm, which subsequently "Old Drum" owned by Charles Burden came onto the farm, and the dog was shot to death by Hornsby. There seems to be little dispute over these facts of the case. Burden sued Hornsby for $150 in damages as to the killing of "Old Drum" (this was the maximum monetary penalty allowed by local law at the time).

During the trial, Vest boastfully vowed outside of court that he would "win the case or apologize to every dog in Missouri," and did indeed ultimately prevail, though the jury awarded just $50 rather than the sought for $150. The case was subsequently appealed and eventually landed at the Missouri Supreme Court, where Vest also prevailed.

Over time, the case and Vest became rather famous for his closing argument in the original trial, gaining fame not especially due to his handling of the case per se but because of his closing remarks. The renown of the closing argument is known for explicitly not having made any reference to the case specifics, nor citing any testimony or evidence presented and served seemingly wholly as a eulogy or tribute about the deceased dog (which, for purposes herein in this paper, showcase the potential significance of the use of sentiment, and the use of opinion, respectively).

The homage has become famously known as the *Tribute to a Dog*.

Here is the closing argument made by Vest (source: https://www.historyplace.com/speeches/vest.htm):

"The best friend a man has in the world may turn against him and become his enemy. His son or daughter that he has reared with loving care may prove ungrateful. Those who are nearest and dearest to us, those whom we trust with our happiness and our good name may become traitors to their faith. The money that a man has, he may lose. It flies away from him, perhaps when he needs it most. A man's reputation may be sacrificed in a moment of ill-considered action. The people who are prone to fall on their knees to do us honor when success is with us, may be the first to throw the stone of malice when failure settles its cloud upon our heads."

"The one absolutely unselfish friend that man can have in this selfish world, the one that never deserts him, the one that never proves ungrateful or treacherous is his dog. A man's dog stands by him in prosperity and in poverty, in health and in sickness. He will sleep on the cold ground, where the wintry winds blow and the snow drives fiercely, if only he may be near his master's side. He will kiss the hand that has no food to offer. He will lick the wounds and sores that come in encounters with the roughness of the world. He guards the sleep of his pauper master as if he were a prince. When all other friends desert, he remains. When riches take wings, and reputation falls to pieces, he is as constant in his love as the sun in its journey through the heavens."

"If fortune drives the master forth, an outcast in the world, friendless and homeless, the faithful dog asks no higher privilege than that of accompanying him, to guard him against danger, to fight against his enemies. And when the last scene of all comes, and death takes his master in its embrace and his body is laid away in the cold ground, no matter if all other friends pursue their way, there by the graveside will the noble dog be found, his head between his paws, his eyes sad, but open in alert watchfulness, faithful and true even in death."

This closing argument has been used as a case study in the use of legal rhetoric and its impact, for which the layers of sentiment and opinion are readily detectible.

**1.4 Computer-based Aided LSAOM**

Efforts to undertake Legal Sentiment Analysis and Legal Opinion Mining have historically been performed by humans, doing so upon inspecting or observing fellow humans, including their real-time behavior, their recorded behavior via video or audio recordings, and their written words via printed or online narratives. An attorney for example might study the efforts of their opponent at trial to try and discern how they are using opinion or sentiment, potentially countering or objecting at an advantageous opportunity to do so, or detected to then seek to deflate the efforts of the opposing counsel. Experts at social psychology might be employed to examine jurors or judges and seek to interpret their sentiments and opinions.

Upon the advent of computers, attempts to make use of computational methods to conduct Legal SA and Legal OM have been thinly aided by the use of computer-based approaches. Simplistic uses of computers can be used to do wordcounts and attempt to ascertain embedded sentiments and opinions, while more complex approaches make use of statistical models. Besides analyzing the written word, voice detection and translation software can also be used, along with the use of facial images captured by cameras and the analysis of the video.

Advances in Artificial Intelligence (AI) involving especially Natural Language Processing (NLP) and Machine Learning (ML) are increasingly bolstering how automation can systematically perform either or both of Sentiment Analysis and Opinion Mining, all of which is being inexorably carried over into engagement within a legal context for improving LSAOM capabilities.

For example, Wyner and Moens [82] make use of a specialized context-free grammar technique to employ NLP for the LSAOM of legal cases: "This paper describes recent approaches using text-mining to automatically profile and extract arguments from legal cases. We outline some of the background context and motivations. We then turn to consider issues related to the construction and composition of corpora of legal cases. We show how a Context-Free Grammar can be used to extract arguments, and how ontologies and Natural Language Processing can identify complex information such as case factors and participant roles. Together the results bring us closer to automatic identification of legal arguments."

In the work by Liu and Chen [55], a two-phased approach of feature extraction from precedents is used to classify judgments per Sentiment Analysis and Opinion Mining: "Factual scenario analysis of a judgment is critical to judges during sentencing. With the increasing number of legal cases, professionals typically endure heavy workloads on a daily basis. Although a few previous studies have applied information technology to legal cases, according to our research, no prior studies have predicted a pending judgment using legal documents. In this article, we introduce an innovative solution to predict relevant rulings. The proposed approach employs text mining methods to extract features from precedents and applies a text classifier to automatically classify judgments according to sentiment analysis. This approach can assist legal experts or litigants in predicting possible judgments. Experimental results from a judgment data set reveal that our approach is a satisfactory method for judgment classification."

The gradual advent of legal e-Discovery has also further spurred progress in LSAOM, including as described in this work by Joshi and Deshpande [51]: "e-Discovery Review is a type of legal service that aims at finding relevant electronically stored information (ESI) in a legal case. This requires manual reviewing of large number of documents by legal analysts, thus involving huge costs. In this paper, we investigate the use of IT, specifically text mining techniques, for improving the efficiency and quality of the e-discovery review service. We employ near duplicate detection and automatic classification techniques that can be used to create coherent groups of documents."

Machine Learning advances and the utilization of Deep Learning via large-scale Artificial Neural Networks (ANN) has also sparked LSAOM, such as this research involving the analysis of legal judgments for criminal cases [18]: "Text mining has become an effective tool for analyzing text documents in automated ways. Conceptually, clustering, classification and searching of legal documents to identify patterns in law corpora are of key interest since it aids law experts and police officers in their analyses. In this paper, we develop a document classification, clustering and search methodology based on neural network technology that helps law enforcement department to manage criminal written judgments more efficiently. In order to maintain a manageable number of independent Chinese keywords, we use term extraction scheme to select top-$n$ keywords with the highest frequency as inputs of the Back-Propagation Network (BPN), and select seven criminal categories as target outputs of it. Related legal documents are automatically trained and tested by pre-trained neural network models. In addition, we use Self Organizing Map (SOM) method to cluster criminal written judgments. The research shows that automatic classification and clustering modules classify and cluster legal documents with a very high accuracy. Finally, the search module which uses the previous results helps users find relevant written judgments of criminal cases."

An interesting variant of LSAOM comes to play when considering the use of legal vocabulary for the general public, as typified by this effort utilizing a three-phase prediction (TPP) algorithm [54]: "Applying text mining techniques to legal issues has been an emerging research topic in recent years. Although a few previous studies focused on assisting professionals in the retrieval of related legal documents, to our knowledge, no previous studies could provide relevant statutes to the general public using problem statements. In this work, we design a text mining based method, the three-phase prediction (TPP) algorithm, which allows the general public to use everyday vocabulary to describe their problems and find pertinent statutes for their cases. The experimental results indicate that our approach can help the general public, who are not familiar with professional legal terms, to acquire relevant statutes more accurately and effectively."

Again, as a gentle reminder, realize that as pointed out in Subsection 1.2., within the AI field, generic SA and generic OM are typically treated as one and the same, in the sense that there is no distinction made between that which is sentiment and that which is opinion. They are oftentimes construed as synonymous and interchangeably used in the AI literature. It is argued herein that within the field of law, there is a bona fide case to be made to consider SA and OM to be distinct in their scope, nature, and focus. Thus, Legal Sentiment Analysis is construed herein as per the definition given earlier in this subsection, and Legal Opinion Mining is construed herein as per the definition given earlier in this subsection.

Beyond a legal context, Sentiment Analysis and Opinion Mining have advanced due to interest in analyzing visual social media, going beyond the written word to examine visual content too [85]: "Social media sentiment analysis (also known as opinion mining) which aims to extract people's opinions, attitudes and emotions from social networks has become a research hotspot. Conventional sentiment analysis concentrates primarily on the textual content. However, multimedia sentiment analysis has begun to receive attention since visual content such as images and videos is becoming a new medium for self-expression in social networks. In order to provide a reference for the researchers in this active area, we give an overview of this topic and describe the algorithms of sentiment analysis and opinion mining for social multimedia. Having conducted a brief review on textual sentiment analysis for social media, we present a comprehensive survey of visual sentiment analysis on the basis of a thorough investigation of the existing literature." This same attention to visual elements is likewise entering into the LSAOM realm.

In seeking to discern Legal Opinion Mining, and to some degree Legal Sentiment Analysis, the study by Conrad and Schilder examined legal blogs that were posted online [20]: "We perform a survey into the scope and utility of opinion mining in legal Weblogs (a.k.a. blawgs). The number of 'blogs' in the legal domain is growing at a rapid pace and many potential applications for opinion detection and monitoring are arising as a result. We summarize current approaches to opinion mining before describing different categories of blawgs and their potential impact on the law and the legal profession. In addition to educating the community on recent developments in the legal blog space, we also conduct some introductory opinion mining trials. We first construct a Weblog test collection containing blog entries that discuss legal search tools. We subsequently examine the performance of a language modeling approach deployed for both subjectivity analysis (i.e., is the text subjective or objective?) and polarity analysis (i.e., is the text affirmative or negative towards its subject?). This work may thus help establish early baselines for these core opinion mining tasks."

Related to the topic of online and social media, those such advances have equally stimulated advancement in generic SA and generic OM, and for which then can be carried into LSAOM. Perhaps the most popular focus of social media for undertaking SA and OM consists of examining tweets, such as this study [5]: "Understanding the behavior of people or a particular user using his comments or tweets in various social media is an advancement of the Sentiment Analysis. Sentiment Analysis or Opinion Mining is used to understand the overall sentiments present in the data collected from various social media. The people have more exposure to the outside world due to the existence of the Internet and Various Social Medias like Twitter, Facebook, Instagram, etc. where they will be sharing their thoughts. Cheap and fast communication has made social media more valuable among the public. Social Media data can be used for various scientific and commercial applications. The combination of Sentiment Analysis and Behavior Analysis made the extraction of needed or useful data more easy and simple for various applications which include character analyzing, Depression Testing etc. Moreover, the behavior analysis will be done based on the text and emoticon sentiment score obtained during the analysis."

Similar kinds of studies have examined the reviews of products, as posted online at sites including Amazon, Facebook, etc., as developed in this study on using SA and OM techniques and technologies [49]: "Sentiment Analysis and Opinion Mining is a most popular field to analyze and find out insights from text data from various sources like Facebook, Twitter, and Amazon, etc. It plays a vital role in enabling the businesses to work actively on improving the business strategy and gain an in-depth insight of the buyer's feedback about their product. It involves computational study of behavior of an individual in terms of his buying interest and then mining his opinions about a company's business entity. This entity can be visualized as an event, individual, blog post or product experience. In this paper, Dataset has taken from Amazon which contains reviews of Camera, Laptops, Mobile phones, tablets, TVs, video surveillance. After preprocessing we applied machine learning algorithms to classify reviews that are positive or negative. This paper concludes that, Machine Learning Techniques gives best results to classify the Products Reviews. Naïve Bayes got accuracy 98.17% and Support Vector machine got accuracy 93.54% for Camera Reviews."

As will be addressed in Section 3, this research paper examines the evolving infusion of AI into Legal Sentiment Analysis and Opinion Mining and proposes an alignment with the Levels of Autonomy (LoA) of AI Legal Reasoning (AILR), plus provides additional insights regarding AI LSAOM in its mechanizations and potential impact to the study of law and the practicing of law.

**1.5 Conventional Legal Contexts for LSAOM**

Aristotle philosophized that the law should be free of passion [48].

There is much focus in the theory of law and the practice of law to presumably excise emotion from the nature of law and the practice of law, such that the law is aimed to be entirely objective, free of subjectivity, dispassionate, and carried by the strength of logic and legal argument [41]. Though this might be a desired arrangement, the reality is that the law and the practice of law are intimately bound into human behavior and therefore subject to human sentiment and to human opinion [79] (both "vacuous" opinion, as it were, if non-factual opinion could be so labeled, and fact-based opinion or ostensibly substantiated opinion).

One of the most visible arenas involving the interest in performing LSAOM consists of jury selection. There is a desire to detect so-called emotional loyalties of jurors, tipping their hands as to their presumed likely proclivities while possibly serving on a jury, as explained by Gobin [41]: "Scientific or Systematic Jury Selection (SJS) originated during the Vietnam War Era and remains targeted by the scientific community as a practice more artistic than factual. But while the motives of the procedure and its status as a recognized science are still controversial, a close examination of its methods provides insight into the hidden pathways of emotional assumption relied upon by jury selectors. Jury consultants who practice SJS usually focus on certain strategic markers: demographic classification, behavioral responses, and psychological attributes. On the surface, these categories create a very satisfying array of options."

And, further by [41]: "Jury instructions–such as those in capital penalty phases, which warn jurors that they 'must not be swayed by mere sentiment, conjecture, sympathy, passion, prejudice, public opinion or public feeling,' urge jurors to abandon any emotional loyalties that they had disclosed in voir dire, and to deliver a verdict they believe to be their impartial best. When jurors are selected through a procedure littered with emotional analysis, however, is it then reasonable to expect an impartial verdict from those selected using a partial process? In other words, can a juror who has been selected based on predictions of his or her emotional responses subsequently be placed in a courtroom and enter judgments devoid of those emotions?"

Walker and Shapiro discuss the overall psychology that permeates trials and thusly embodies a cauldron of sentiment and opinion, notwithstanding efforts to keep such "subjective" emulsions at bay [79]: "Despite skepticism that psychologists were mind readers and could manipulate people, coupled with concern over attorneys and even mental health professionals that might overstep their bounds, the area of trial consultation and jury selection has become an important area of forensic psychology. Much of the research and development in this area stems from social psychology, and methods such as public opinion polls, focus groups, mock trials, and analogue jury studies are used to accomplish the goals of, preparing witnesses for their statements and selecting (or, rather, deselecting) jurors in the *voir dire*, or the process by which juries are chosen for a trial. Forensic psychologists serving as trial consultants also use research to assist the attorney in trial strategies such as decisions on which pieces of evidence to emphasize, how to arrange evidence in terms of order of presentation, preparation of opening and closing statements, and determining when and if a change of venue is necessary in order to obtain a fair trial for clients, among other tasks."

Judges are also subject to sentiment and opinion, and the psychology of trial judging showcases the significant impacts therein [77]: "Trial court judges play a crucial role in the administration of justice for both criminal and civil matters. Although psychologists have studied juries for many decades, they have given relatively little attention to judges. Recent writings, however, suggest increasing interest in the psychology of judicial decision making. This essay reviews several selected topics where judicial discretion appears to be influenced by psychological dispositions, but cautions that a mature psychology of judging field will need to consider the influence of the bureaucratic court setting in which judges are embedded, their legal training, and the constraints of legal precedent."

It might be assumed that judges would be able to readily overcome sentiment, and yet some studies have indicated that even when directly informed to exclude biasing material, they appeared (along with jurors) to nonetheless not be able to logically and "objectively" do so [53]: "Reviews the presumptions and the differential treatment accorded American judges and jurors by the civil procedure system. An experiment was conducted in which 88 judges and 104 jurors were exposed to potentially biasing material with respect to a civil trial vignette. Judges and jurors randomly received 1 of 3 versions of a product liability case: no exposure to biasing material, exposure with a judicial decision to exclude the material, and exposure with a judicial decision to admit the material. Ss were asked to indicate (1) whether they would find the defendant liable or not liable and (2) their level of confidence in their decisions. Results suggest that judges and jurors may be similarly influenced by such exposure, regardless of whether the biasing material was ruled admissible or inadmissible."

Importantly, rather than taking a blind eye toward the inclusion of sentiment, some argue that it is better to put the sentiment at front and center, perhaps even asserting that the embodiment of emotions and feeling by judges can be considered a useful element for rendering good legal judgments [48]: "There has been an explosion of emotion research in which emotions are no longer seen in opposition to reason. Instead, emotions are increasingly appreciated as being indispensable in cognitive processes because they comprise sets of perceptions and evaluations that enable judgment. Emotions are no longer set aside as mere obstacles to good judgment. At the same time, however, it is a well-known fact that emotions also often prevent people from judging carefully."

And continuing [48]: "The question is raised, for example, whether it is desirable for judges to express their emotions and what the right way would be for them to do so. Questions like this introduce new arguments into the ongoing debate in legal philosophy about legal positivism: about its rationalist ideal of the dispassionate judge who merely applies rules."

**1.6 Hierarchical Nature of LSAOM**

A topic that will be addressed in Section 3 and for which is worthwhile to first introduce in Section 1 consists of the granularity associated with undertaking LSAOM.

It is important to consider the granularity at which a Legal Sentiment Analysis might be undertaken, and likewise at which a Legal Opinion Mining might be undertaken.

Use the *Tribute of a Dog* as a vehicle for exploring the granularity facets of LSAOM. One could examine perhaps the first sentence: "The best friend a man has in the world may turn against him and become his enemy." When the sentence was uttered by Vest, presumably a SA could be done as to his tonality and manner in which he expressed the sentence. From an OM perspective, the sentence could be examined for its essence of opinion expressed, and whether it appeared to be fact-based or non-factually based.

Certainly, we might though desire to inspect the entire initial paragraph, rather than merely the first sentence. Or, we might wish to examine the entire closing argument.

Any of these could be a proper or appropriate granularity at which to undertake an LSAOM or could be inappropriate and produce a false or misleading conclusion arising from the SA and the OM derivations, and thus the application of LSAOM has to be weighed with respect to its value and utility for what amount of granularity it is being applied.

As indicated by Do et al [22]:

"A current research focus for sentiment analysis is the improvement of granularity at aspect level, representing two distinct aims: aspect extraction and sentiment classification of product reviews and sentiment classification of target-dependent tweets. *Deep learning* approaches have emerged as a prospect for achieving these aims with their ability to capture both syntactic and semantic features of text without requirements for high-level feature engineering, as is the case in earlier methods."

Similarly, in the work by Gamal et al [38]:

"SA is categorized into three main levels—the aspect or feature level (AL), the sentence level (SL) and the document level (DL). The AL refers to classify the sentiments that are expressed on various features or aspects of an entity. In the SL, the fundamental concern is to pick whether each sentence infers a positive, negative or neutral opinion. In the DL, the basic concern is to classify whether the whole opinion in a document implies a positive or negative sentiment. The SL and DL analyses are insufficient to precisely monitor what people accept and reject. This research focuses on the document level of sentiment analysis."

Overall, those advancing the theory of law in the realm of LSAOM, and those practicing law by the leveraging of LSAOM, need to be fully aware of the granularity facets, likely of a hierarchical and at times circular nature in any particular legal case or legal contextual application (this is further discussed in Section 3).

## 2 Levels of Autonomy (LOA) of AI Legal Reasoning (AILR)

In this section, a framework for the autonomous levels of AI Legal Reasoning is summarized and is based on the research described in detail in Eliot [28] [29] [30] [31] [32] [33] [34] [35] [36].

These autonomous levels will be portrayed in a grid that aligns with key elements of autonomy and as matched to AI Legal Reasoning. Providing this context will be useful to the later sections of this paper and will be utilized accordingly.

The autonomous levels of AI Legal Reasoning are as follows:

Level 0: No Automation for AI Legal Reasoning

Level 1: Simple Assistance Automation for AI Legal Reasoning

Level 2: Advanced Assistance Automation for AI Legal Reasoning

Level 3: Semi-Autonomous Automation for AI Legal Reasoning

Level 4: Domain Autonomous for AI Legal Reasoning

Level 5: Fully Autonomous for AI Legal Reasoning

Level 6: Superhuman Autonomous for AI Legal Reasoning

### 2.1 Details of the LoA AILR

See **Figure A-1** for an overview chart showcasing the autonomous levels of AI Legal Reasoning as via columns denoting each of the respective levels.

See **Figure A-2** for an overview chart similar to Figure A-1 which alternatively is indicative of the autonomous levels of AI Legal Reasoning via the rows as depicting the respective levels (this is simply a reformatting of Figure A-1, doing so to aid in illuminating this variant perspective, but does not introduce any new facets or alterations from the contents as already shown in Figure A-1).

#### 2.1.1 Level 0: No Automation for AI Legal Reasoning

Level 0 is considered the no automation level. Legal reasoning is carried out via manual methods and principally occurs via paper-based methods.

This level is allowed some leeway in that the use of say a simple handheld calculator or perhaps the use of a fax machine could be allowed or included within this Level 0, though strictly speaking it could be said that any form whatsoever of automation is to be excluded from this level.

#### 2.1.2 Level 1: Simple Assistance Automation for AI Legal Reasoning

Level 1 consists of simple assistance automation for AI legal reasoning.

Examples of this category encompassing simple automation would include the use of everyday computer-based word processing, the use of everyday computer-based spreadsheets, the access to online legal documents that are stored and retrieved electronically, and so on.

By-and-large, today's use of computers for legal activities is predominantly within Level 1. It is assumed and expected that over time, the pervasiveness of automation will continue to deepen and widen, and eventually lead to legal activities being supported and within Level 2, rather than Level 1.

#### 2.1.3 Level 2: Advanced Assistance Automation for AI Legal Reasoning

Level 2 consists of advanced assistance automation for AI legal reasoning.

Examples of this notion encompassing advanced automation would include the use of query-style Natural Language Processing (NLP), Machine Learning (ML) for case predictions, and so on.

Gradually, over time, it is expected that computer-based systems for legal activities will increasingly make use of advanced automation. Law industry technology that was once at a Level 1 will likely be refined, upgraded, or expanded to include advanced capabilities, and thus be reclassified into Level 2.

#### 2.1.4 Level 3: Semi-Autonomous Automation for AI Legal Reasoning

Level 3 consists of semi-autonomous automation for AI legal reasoning.

Examples of this notion encompassing semi-autonomous automation would include the use of Knowledge-Based Systems (KBS) for legal reasoning, the use of Machine Learning and Deep Learning (ML/DL) for legal reasoning, and so on.

Today, such automation tends to exist in research efforts or prototypes and pilot systems, along with some commercial legal technology that has been infusing these capabilities too.

### 2.1.5 Level 4: Domain Autonomous for AI Legal Reasoning

Level 4 consists of domain autonomous computer-based systems for AI legal reasoning.

This level reuses the conceptual notion of Operational Design Domains (ODDs) as utilized in the autonomous vehicles and self-driving cars levels of autonomy, though in this use case it is being applied to the legal domain [24] [25] [26] [27]. Essentially, this entails any AI legal reasoning capacities that can operate autonomously, entirely so, but that is only able to do so in some limited or constrained legal domain.

### 2.1.6 Level 5: Fully Autonomous for AI Legal Reasoning

Level 5 consists of fully autonomous computer-based systems for AI legal reasoning.

In a sense, Level 5 is the superset of Level 4 in terms of encompassing all possible domains as per however so defined ultimately for Level 4. The only constraint, as it were, consists of the facet that the Level 4 and Level 5 are concerning human intelligence and the capacities thereof. This is an important emphasis due to attempting to distinguish Level 5 from Level 6 (as will be discussed in the next subsection)

It is conceivable that someday there might be a fully autonomous AI legal reasoning capability, one that encompasses all of the law in all foreseeable ways, though this is quite a tall order and remains quite aspirational without a clear cut path of how this might one day be achieved. Nonetheless, it seems to be within the extended realm of possibilities, which is worthwhile to mention in relative terms to Level 6.

### 2.1.7 Level 6: Superhuman Autonomous for AI Legal Reasoning

Level 6 consists of superhuman autonomous computer-based systems for AI legal reasoning.

In a sense, Level 6 is the entirety of Level 5 and adds something beyond that in a manner that is currently ill-defined and perhaps (some would argue) as yet unknowable. The notion is that AI might ultimately exceed human intelligence, rising to become superhuman, and if so, we do not yet have any viable indication of what that superhuman intelligence consists of and nor what kind of thinking it would somehow be able to undertake.

Whether a Level 6 is ever attainable is reliant upon whether superhuman AI is ever attainable, and thus, at this time, this stands as a placeholder for that which might never occur. In any case, having such a placeholder provides a semblance of completeness, doing so without necessarily legitimatizing that superhuman AI is going to be achieved or not. No such claim or dispute is undertaken within this framework.

## 3 LSAOM and LoA AILR

In this Section 3, various aspects of Legal Sentiment Analysis and Opinion Mining (LSAOM) will be identified and discussed with respect to AI Legal Reasoning (AILR). A series of diagrams and illustrations are included to aid in depicting the points being made. In addition, the material draws upon the background and LSAOM research literature indicated in Section 1 and combines with the material outlined in Section 2 on the Levels of Autonomy (LoA) of AI Legal Reasoning.

### 3.1 LSAOM Aligned with LoA AILR

The nature and capabilities of Legal Sentiment Analysis and Opinion Mining will vary across the Levels of Autonomy for AI Legal Reasoning. Though it is argued in this paper that legal-oriented SA and legal-oriented OM are two distinct facets, which are often intertwined but not necessarily so, and for which they are most decidedly not considered as synonymous with each other, they nonetheless can be treated as two akin capacities that will likely advance and mature correspondingly in the same overarching manner, over time and amidst the advent of AILR levels of autonomy.

Refer to **Figure B-1**.

As indicated, Legal Sentiment Analysis and Opinion Mining becomes increasingly more sophisticated and advanced as the AI Legal Reasoning increases in capability. To aid in typifying the differences between each of the Levels in terms of the incremental advancement of LSAOM, the following phrasing is used:

- Level 0: **n/a**
- Level 1: **Rudimentary Detection**
- Level 2: **Complex Detection**
- Level 3: **Symbolic Intertwined**
- Level 4: **Domain Perceptive**
- Level 5: **Holistic Perceptive**
- Level 6: **Pansophic Perceptive**

Briefly, each of the levels of LSAOM is described next.

At Level 0, there is an indication of "n/a" at Level 0 since there is no AI capability at Level 0 (the *No Automation* level of the LoA).

At Level 1, the LoA is *Simple Assistance Automation* and this can be used to undertake Legal Sentiment Analysis and Opinion Mining though it is rated or categorized as being rudimentary and making use of relatively simplistic calculative models and formulas. Thus, this is coined as "Rudimentary Detection."

At Level 2, the LoA is *Advanced Assistance Automation* and the LSAOM is coined as "Complex Detection," which is indicative of Legal Sentiment Analysis and Opinion Mining being performed in a more advanced manner than at Level 1. This consists of complex statistical methods such as those techniques mentioned in Section 1 of this paper. To date, most of the research and practical use of Legal Sentiment Analysis and Opinion Mining has been within Level 2. Future efforts are aiming at Level 3 and above.

At Level 3, the LoA is *Semi-Autonomous Automation* and the LSAOM is coined as "Symbolic Intermixed," which can undertake Legal Sentiment Analysis and Opinion Mining at an even more advanced capacity than at Level 2. Recall, in Level 2, the focus tended to be on traditional numerical formulations for LSAOM, albeit sophisticated in the use of statistical models. In Level 3, the symbolic capability is added and fostered, including at times acting in a hybrid mode with the conventional numerical and statistical models. Generally, the work at Level 3 to-date has primarily been experimental, making use of exploratory prototypes or pilot efforts.

At Level 4, the LoA is *AILR Domain Autonomous* and the LSAOM coined as "Domain Perceptive," meaning that this can be used to perform Legal Sentiment Analysis and Opinion Mining within particular specialties of domains or subdomains of the legal field, but does not necessarily cut across the various domains and is not intended to be able to do so. The capacity is done in a highly advanced manner, incorporating the Level 3 capabilities, along with exceeding those levels and providing a more fluent and capable perceptive means.

At Level 5, the LoA is *AILR Fully Autonomous,* and the LSAOM coined as "Holistic Perceptive," meaning that the use of Legal Sentiment Analysis and Opinion Mining can go across all domains and subdomains of the legal field. The capacity is done in a highly advanced manner, incorporating the Level 4 capabilities, along with exceeding those levels and providing a more fluent and capable perceptive means.

At Level 6, the LoA is *AILR Superhuman Autonomous*, which as a reminder from Section 2 is not a capability that exists and might not exist, though it is included as a provision in case such a capability is ever achieved. In any case, the LSAOM at this level is considered "Pansophic Perceptive" and would encapsulate the Level 5 capabilities, and then go beyond that in a manner that would leverage the AI superhuman capacity.

**3.2 Framework of Legal Sentiment Analysis and Opinion Mining**

Based on the discussion in Section 1, it is useful to consider the overarching nature of the approaches utilized in ascertaining Legal Sentiment Analysis and Opinion Mining and provide a framework for establishing the elements involved.

Refer to **Figure B-2**.

The framework indicates that LSAOM consists of two facets that are often intertwined, though can be distinctively articulated as they proffer differing scope, nature, and focus. Sentiment Analysis (SA) is utilized for the detection of expressed or implied sentiment about a legal matter within the context of a legal milieu, while Opinion Mining (OM) is utilized for the identification and illumination of explicit or implicit opinion accompaniments immersed within legal discourse.

For the Legal Sentiment Analysis, three major elements are consisting of: (1) SA Visual, (2) SA Oral, and (3) SA Written. The SA Visual is typically aimed at facial recognition for sentiment detection, but other visual indications can be encompassed, such as body language, posturing, etc. SA Oral entails discourse that is orally expressed rather than in writing. Within SA Oral, there is the tonality that is examined to aid in ascertaining the sentiment expression, along with the words spoken as part of a legal narrative. SA Written entails discourse that consists of written narrative.

For the Legal Opinion Mining, there are two major elements: (1) OM Oral, and (2) OM Written. Underlying each of these two elements is the vital aspect of detecting whether an expressed or implied opinion is seemingly facts-based or whether it is non-factual based.

Refer next to **Figure B-3**.

Customarily, the use of Legal Sentiment Analysis and Opinion Mining entails examining human utterances and expressions. This might consist of the sentiment and/or opinions of a judge, of a jury member, of an attorney, and so on.

In the future, in addition to the use of LSAOM on human utterances and expressions, it is anticipated that the LSAOM will be applied to AI utterances and expressions. This might seem farfetched at this time, and yet the future might indeed involve AI systems that will be providing legal arguments and undertaking legal discourses, which today is rudimentary and minimal at best by any existent AILIR.

Consider the four-square grid in **Figure B-4**.

As shown in Figure B-4, consider a four-square grid that presents some notable nuances between the Legal Sentiment Analysis and the Legal Opinion Mining capabilities.

On the vertical axis are the two capacities, Legal Sentiment Analysis, and Legal Opinion Mining, respectively. Along the horizontal axis is time as divided into real-time versus offline.

Generally, the primary use of Legal Sentiment Analysis occurs when analyzing in real-time the facial and voice aspects of a person speaking (or, an AI system speaking). This can be undertaken to gauge a "now expressing" detection of sentiment. Legal Sentiment Analysis can also be used in an offline mode, such as analyzing an audio or video recording, or for examining a written narrative.

Generally, the primary use of Legal Opinion Mining occurs when analyzing an offline transcript of written words (this could be in an audio or video format and transformed into a written word format). This is utilized to ascertain explicitly or implicitly expressed opinions, along with whether they are fact-based or non-factual based (as within the context of the provided narrative being used as the scope of the analysis). Legal Opinion Mining can also be used in real-time settings, though this is a usually less impactful manner.

In practice, the Legal Sentiment Analysis might not be able to ascertain any semblance of sentiment expressed. Thus, Legal Sentiment Analysis can consist of two polarity states, SA SD1 (detected) and SA SD0 (not detected).

Likewise, in Legal Opinion Mining, an opinion might not be detected (coding of OM OD0), or might be detected and be factual based (coding of OM OD1-FB) or non-factual based (coding of OM OD1-NFB).

Next, refer to **Figure B-5**.

Earlier in this subsection, it was mentioned that the LSAOM can be applied to AI utterances and expressions, seeking to detect sentiments and opinions as exhibited by an AI system. This does not suggest that the AI is sentient, and leaves aside the question about the potential of AI becoming sentient. In short, an AI system could present a legal argument or legal discourse that embodies sentiment and opinion, doing so without any need per se of somehow having reached sentience. This possibility can occur by a certain kind of happenstance in how the AI has been set up and been designed.

In a somewhat similar vein, the LSAOM capabilities can be used to generate sentiments and opinions, doing so by an AI "reverse" generating approach. Thus, the point being that the same capacity of detection can also potentially be "reversed" into becoming generators.

Refer to **Figure B-6**.

Another important facet of Legal Sentiment Analysis and Opinion Mining is the locus of granularity.

A semblance of sentiment can be sought at a macro-level. Also, sentiment can be sought at a micro-level, and also at a sub-micro-level, and so on. These can be distinctive at each such level or can be rolled-up or rolled-down. Similarly, detection of opinion can be sought at a macro-level. In addition, an opinion can be sought at a micro-level, and also at a sub-micro-level, and so on. These can be distinctive at each such level or can be rolled-up or rolled-down.

This is an important facet of LSAOM, namely that it is unlikely that any given SA or any given OM will be upon a monolith that exhibits one and only one such sentiment or opinion. Instead, the greater likelihood is that sentiment will differ at various levels of expression and opinion will differ at various levels of expression.

**4 Additional Considerations and Future Research**

As earlier indicated, efforts to undertake Legal Sentiment Analysis and Opinion Mining have historically been performed by human hand and cognition, and only thinly aided in more recent times by the use of computer-based approaches. Advances in Artificial Intelligence (AI) involving especially Natural Language Processing (NLP) and Machine Learning (ML) are increasingly bolstering how automation can systematically perform either or both of Sentiment Analysis and Opinion Mining, all of which is being inexorably carried over into engagement within a legal context for improving LSAOM capabilities. This research paper has examined the evolving infusion of AI into Legal Sentiment Analysis and Opinion Mining and proposed alignment with the Levels of Autonomy (LoA) of AI Legal Reasoning (AILR), plus provided additional insights regarding AI LSAOM in its mechanizations and potential impact to the study of law and the practicing of law.

Artificial Intelligence (AI) based approaches have been increasingly utilized and will undoubtedly have a pronounced impact on how LSAOM is performed and its use in the practice of law, which will inevitably also have an impact upon theories of the law.

Future research is needed to explore in greater detail the manner and means by which AI-enablement will occur in the law along with the potential for both positive and adverse consequences due to LSAOM. Autonomous AILR is likely to materially impact the effort, theory, and practice of Legal Sentiment Analysis and Opinion Mining, including as a minimum playing a notable or possibly even pivotal role in such advancements.

**About the Author**

Dr. Lance Eliot is the Chief AI Scientist at Techbrium Inc. and a Stanford Fellow at Stanford University in the CodeX: Center for Legal Informatics. He previously was a professor at the University of Southern California (USC) where he headed a multi-disciplinary and pioneering AI research lab. Dr. Eliot is globally recognized for his expertise in AI and is the author of highly ranked AI books and columns.

**Figure A-1**

| Level | Descriptor | Examples | Automation | Status |
|---|---|---|---|---|
| | | **AI & Law: Levels of Autonomy For AI Legal Reasoning (AILR)** | | |
| 0 | No Automation | Manual, paper-based (no automation) | None | De Facto - In Use |
| 1 | Simple Assistance Automation | Word Processing, XLS, online legal docs, etc. | Legal Assist | Widely In Use |
| 2 | Advanced Assistance Automation | Query-style NLP, ML for case prediction, etc. | Legal Assist | Some In Use |
| 3 | Semi-Autonomous Automation | KBS & ML/DL for legal reasoning & analysis, etc. | Legal Assist | Primarily Prototypes & Research Based |
| 4 | AILR Domain Autonomous | Versed only in a specific legal domain | Legal Advisor (law fluent) | None As Yet |
| 5 | AILR Fully Autonomous | Versatile within and across all legal domains | Legal Advisor (law fluent) | None As Yet |
| 6 | AILR Superhuman Autonomous | Exceeds human-based legal reasoning | Supra Legal Advisor | Indeterminate |

*Figure 1: AI & Law - Autonomous Levels by Rows*  Source Author: Dr. Lance B. Eliot  V1.3

**Figure A-2**

### AI & Law: Levels of Autonomy For AI Legal Reasoning (AILR)

| | Level 0 | Level 1 | Level 2 | Level 3 | Level 4 | Level 5 | Level 6 |
|---|---|---|---|---|---|---|---|
| **Descriptor** | No Automation | Simple Assistance Automation | Advanced Assistance Automation | Semi-Autonomous Automation | AILR Domain Autonomous | AILR Fully Autonomous | AILR Superhuman Autonomous |
| **Examples** | Manual, paper-based (no automation) | Word Processing, XLS, online legal docs, etc. | Query-style NLP, ML for case prediction, etc. | KBS & ML/DL for legal reasoning & analysis, etc. | Versed only in a specific legal domain | Versatile within and across all legal domains | Exceeds human-based legal reasoning |
| **Automation** | None | Legal Assist | Legal Assist | Legal Assist | Legal Advisor (law fluent) | Legal Advisor (law fluent) | Supra Legal Advisor |
| **Status** | De Facto – In Use | Widely In Use | Some In Use | Primarily Prototypes & Research-based | None As Yet | None As Yet | Indeterminate |

Figure 2: AI & Law - Autonomous Levels by Columns        Source Author: Dr. Lance B. Eliot

V1.3

**Figure B-1**

| | Level 0 | Level 1 | Level 2 | Level 3 | Level 4 | Level 5 | Level 6 |
|---|---|---|---|---|---|---|---|
| | | | | | | | |
| **Descriptor** | No Automation | Simple Assistance Automation | Advanced Assistance Automation | Semi-Autonomous Automation | AILR Domain Autonomous | AILR Fully Autonomous | AILR Superhuman Autonomous |
| **Examples** | Manual, paper-based (no automation) | Word Processing, XLS, online legal docs, etc. | Query-style NLP, ML for case prediction, etc. | KBS & ML/DL for legal reasoning & analysis, etc. | Versed only in a specific legal domain | Versatile within and across all legal domains | Exceeds human-based legal reasoning |
| **Automation** | None | Legal Assist | Legal Assist | Legal Assist | Legal Advisor (law fluent) | Legal Advisor (law fluent) | Supra Legal Advisor |
| **Status** | De Facto – In Use | Widely In Use | Some In Use | Primarily Prototypes & Research-based | None As Yet | None As Yet | Indeterminate |
| **AI-Enabled Legal Sentiment Analysis/OM** | n/a | Rudimentary Detection | Complex Detection | Symbolic Intermixed | Domain Perceptive | Holistic Perceptive | Pansophic Perceptive |

Legal Sentiment Analysis/OM: Levels of Autonomy For AI Legal Reasoning (AILR)

*Figure 1: Legal Sentiment Analysis/OM - Autonomous Levels of AILR by Columns*  Source Author: Dr. Lance B. Eliot

V1.3

**Figure B-2**

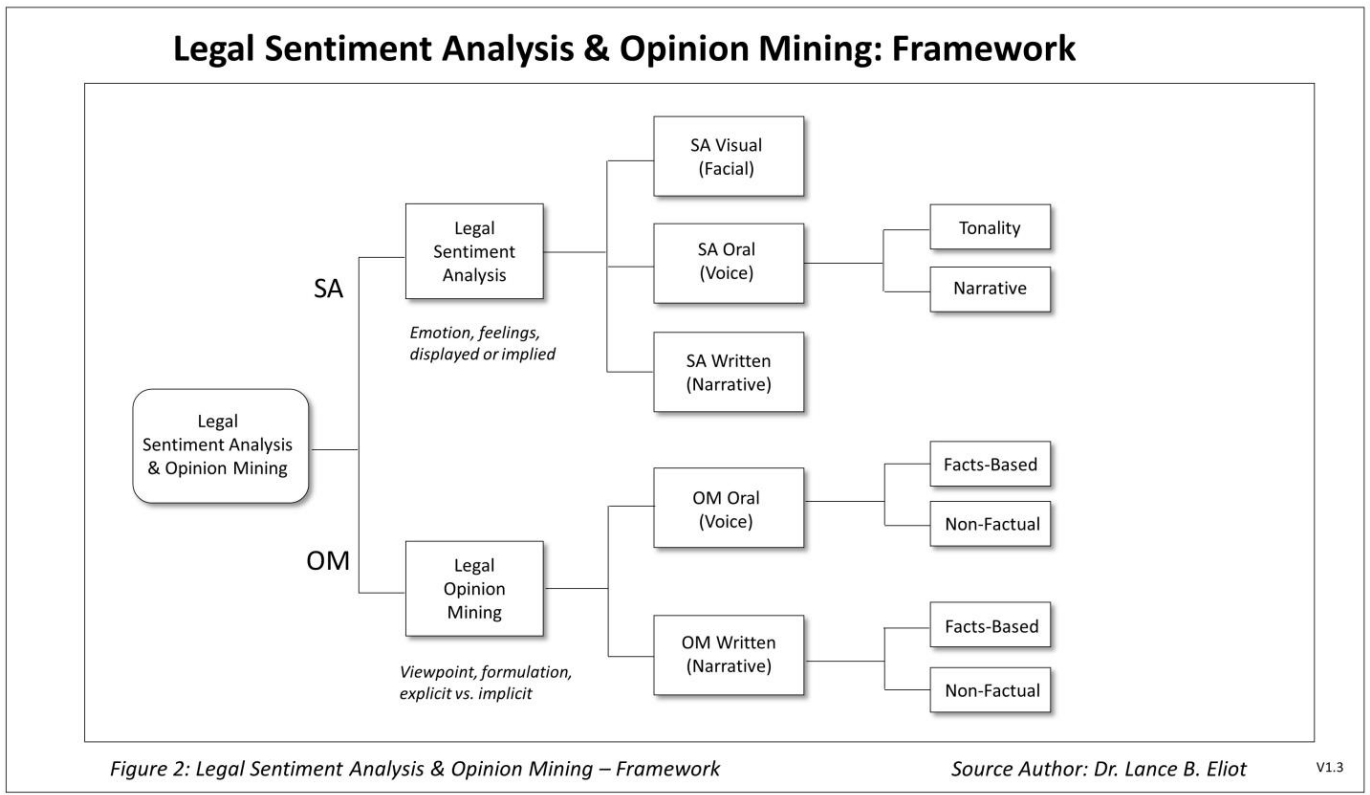

Figure 2: Legal Sentiment Analysis & Opinion Mining – Framework   Source Author: Dr. Lance B. Eliot

**Figure B-3**

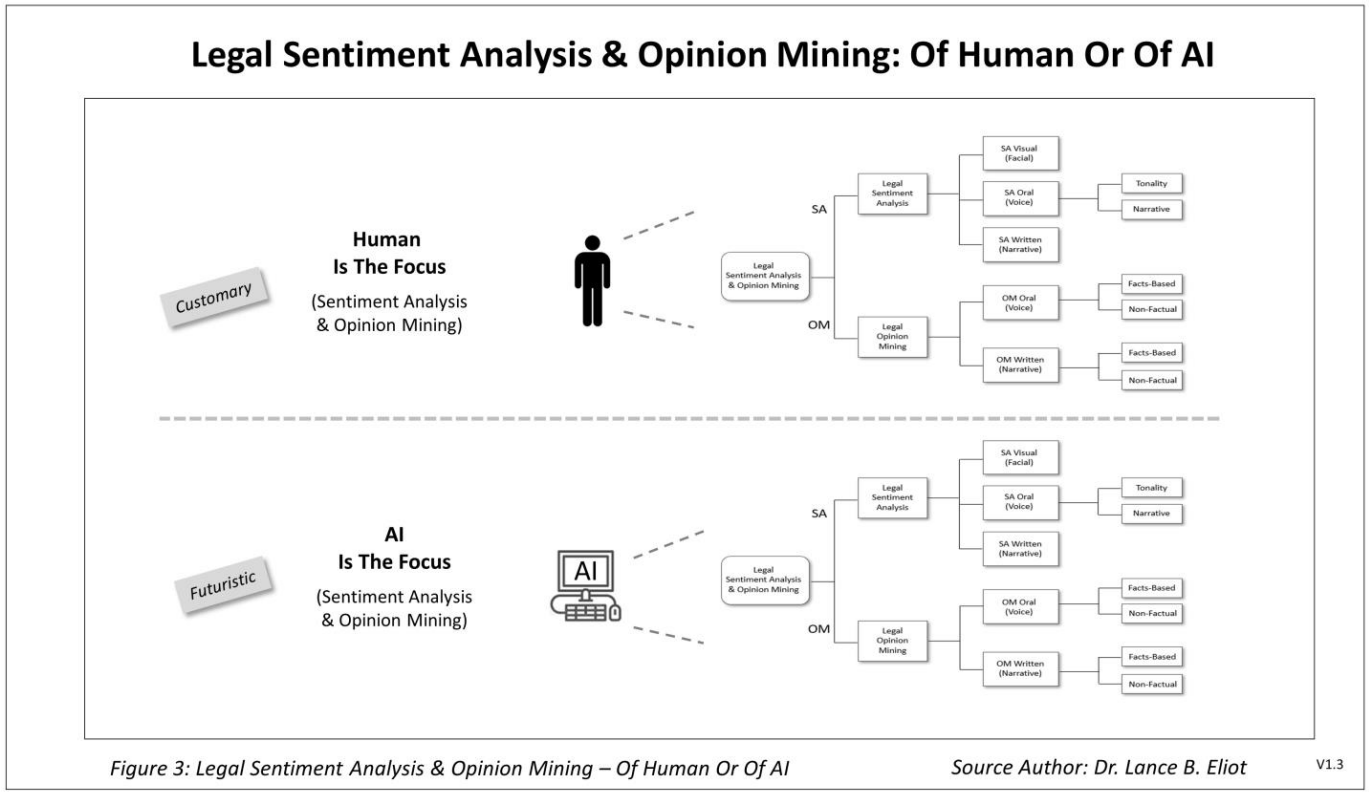

**Figure B-4**

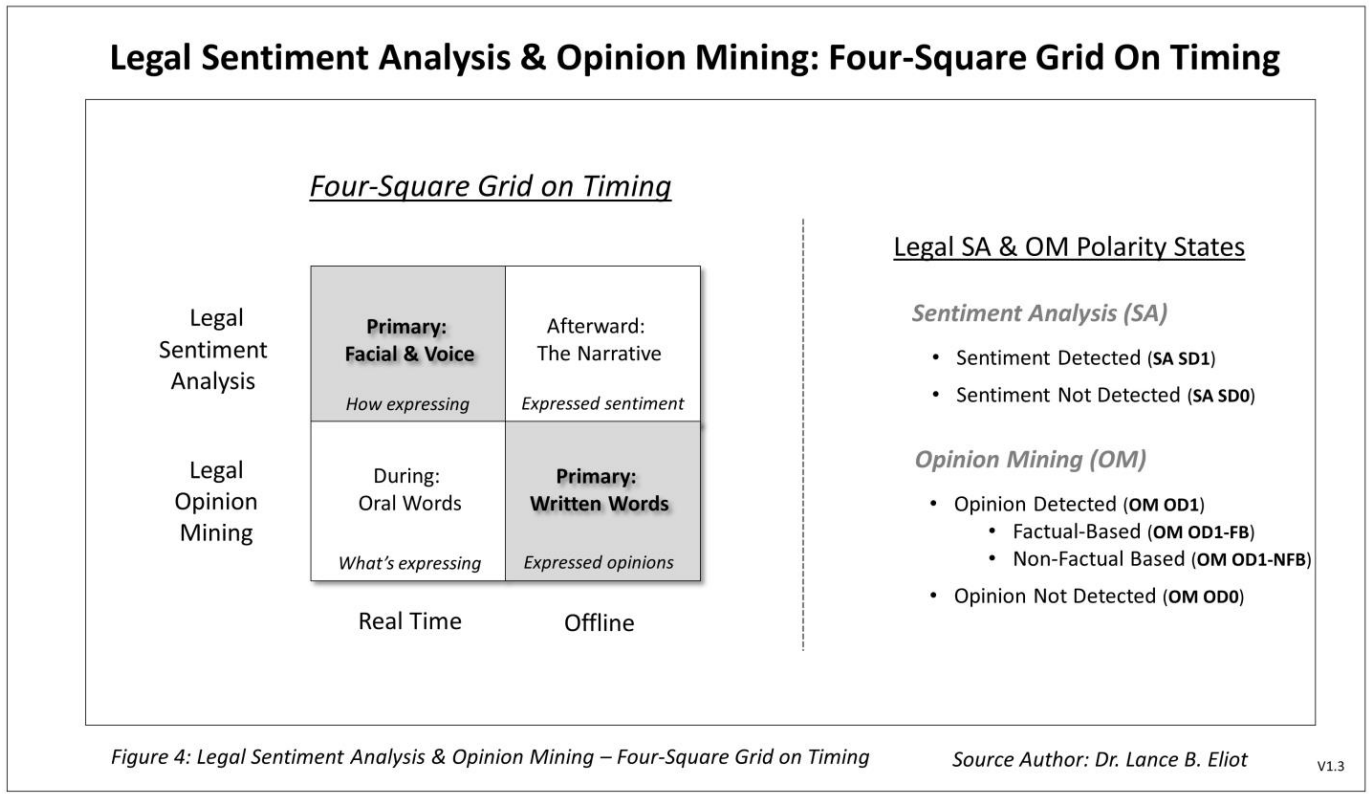

Figure 4: Legal Sentiment Analysis & Opinion Mining – Four-Square Grid on Timing   Source Author: Dr. Lance B. Eliot

**Figure B-5**

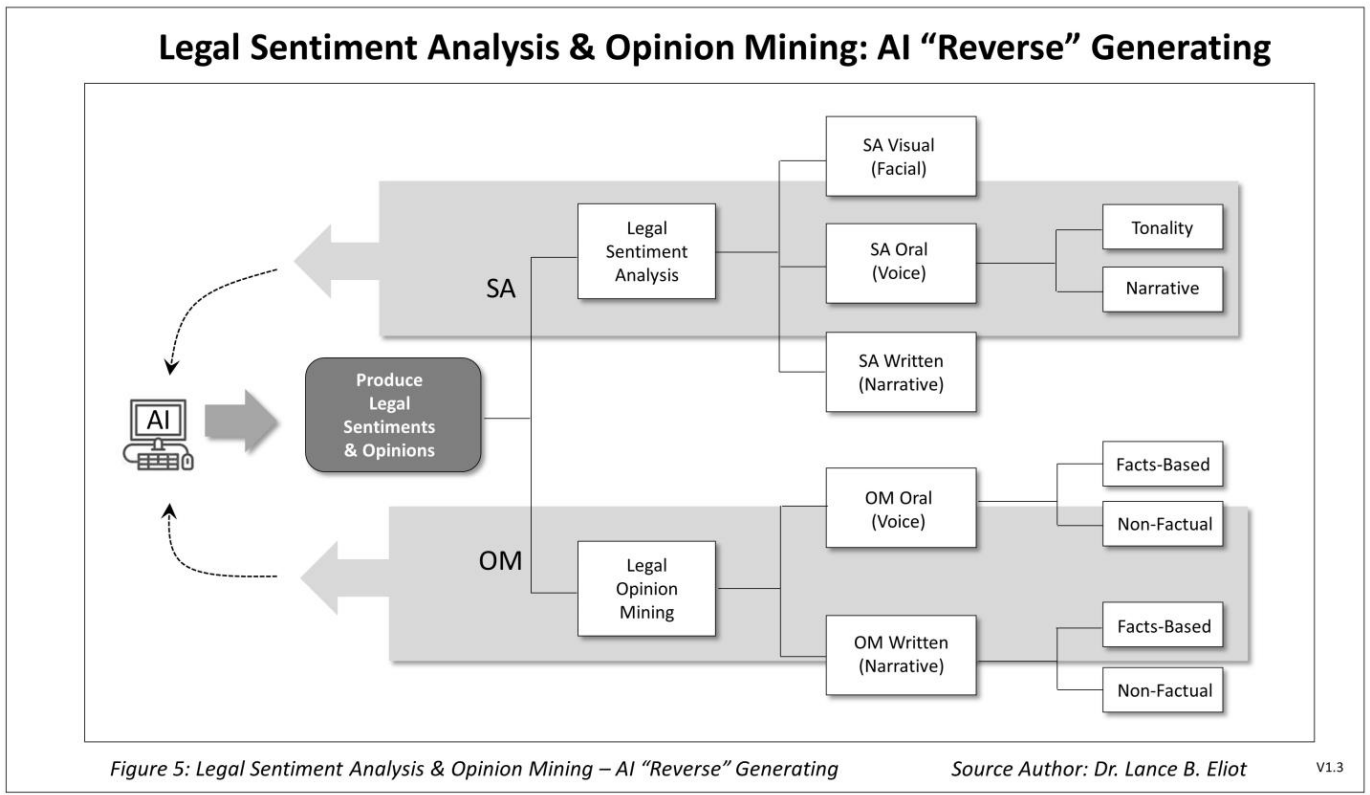

**Figure B-6**

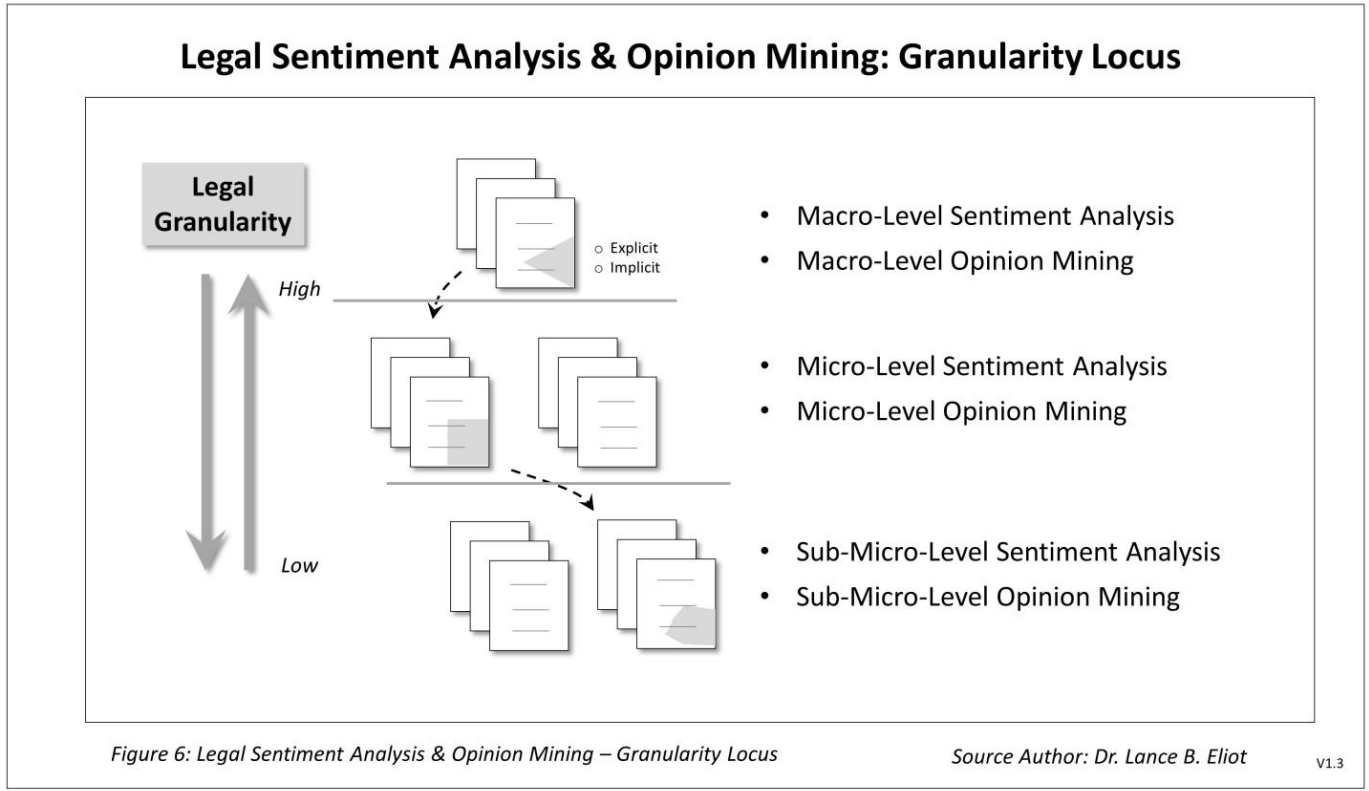

Figure 6: Legal Sentiment Analysis & Opinion Mining – Granularity Locus